\documentstyle{article}
\begin{document}
\title{
Initial-Final State Subspace of the SU(n) Gauge Theory with Explicit Gauge
Field Mass Term}
\author{Ze-sen Yang, Xianhui Li and Xiaolin Chen \\
Department of Physics,  Peking University, Beijing 100871, CHINA }
\date{\today}
\maketitle
\begin{abstract}
     As a part of our study on the SU(n) gauge theory with explicit gauge
field mass term this paper is devoted to form the Gupta-Bleuler
subspace of the initial-final states in the scattering process.
\end{abstract}
\begin{center}
\end{center}
\vspace{3mm}
\par
    We proved in Ref. [1] that the explicit gauge field term is harmless to
the renormalisability of the SU(n) gauge theory. Furthermore we developed
a method in in Ref. [2] to construct the scattering matrix with the
renormalized field functions as basical variables. In this paper we will
show how to form the Gupta-Bleuler subspace of the initial-final states in
the scattering process.
\par
    As was pointed out in section 2, the particles in the initial or final
states are described by the following Lagrangian and Lorentz condition:
\begin{eqnarray}
&& {\cal L}_0 = -\frac{1}{4} F^{[0]}_{a \mu \nu} F_a^{[0]\mu \nu}
  + \frac{1}{2} M^2 A_{a \mu}A_a^{\mu} + {\cal L}_{\psi}
  - \big( \partial_{\mu} \overline{\omega}_a(x) \big)
  \partial^{\mu} \omega_a(x) \,, \\
&& \partial^{\mu} A_{a \mu} = 0 \,,
\end{eqnarray}
where $A_{a \mu}, \omega_a,\overline{\omega}_a$ stand for the renormalized
gauge and ghost field functions. $M$ is the renormalized mass parameter of
the gauge fields. $F^{[0]}_{a \mu \nu}$ is defined by
\begin{eqnarray}
&& F^{[0]}_{a\mu\nu}(x)
    = \partial_{\mu} A_{a\nu}(x) - \partial_{\nu} A_{a\mu}(x) \,.
\end{eqnarray}
According to the operator description of Ref. [2] we should keep the ghost
term in equation (1) and form the initial-final state subspace by following
the Gupta-Bleuler quantization method. To this end we first use the Lorentz
and modify (1) into
\begin{eqnarray}
\widetilde{{\cal L}}_0(x) = {\cal L}_{GB}(x)
+ \frac{1}{2} M^2 A_{a \mu}(x) A_{a}^{\mu}(x) + {\cal L}_{\psi}(x)
- \big(\partial_{\mu}\overline{\omega}_a(x)\big)\partial^{\mu}\omega_a(x)\,,
\end{eqnarray}
where
${\cal L}_{GB}(x)$ is a Gupta--Bleuler Lagrangian
$$
{\cal L}_{GB}(x)
       =  -\frac{1}{4} F^{[0]}_{a \mu \nu} F_a^{[0]\mu \nu}
         - \frac{1}{2} \big( \partial^{\mu} A_{a \mu} \big)^2 \,.
$$
Then we disregard the Lorentz condition and perform a canonical
quantization with $\widetilde{{\cal L}}_0(x)$ and define the
corresponding Hamiltonian operator $H_0$ and field operators.
Moreover we have to choose the so-called physical states and
define a physical initial-final state subspace with the help of
the Lorentz condition and the residual symmetry. We certainly
expect the ghost particles to be excluded from appearing in the
initial-final states.
\par
    Assume that the operators of $\omega_a(x)$ are anti-hermitian and
therefore the operators of $\overline{\omega}_a(x)$ is hermitian.
Let $\omega^{(1)}_a(x)$, $\omega^{(2)}_a(x)$ stand for
$\overline{\omega}_a(x)$, $i\omega_a(x)$ and $\Pi^{(1)}_a$,
$\Pi^{(2)}_a$, $\Pi_{a\mu}$ stand for the canonical momenta
conjugate to $\omega^{(1)}_a(x)$, $\omega^{(2)}_a(x)$, $A_a^{\mu}$
respectively. Thus the operators of these quantities in the
Schrodinger picture are
\begin{eqnarray}
&& A_{a\mu}({\bf x}) = \int \frac{d^3k}{\sqrt{2\Omega(2\pi)^3}}
        \Big[ a_{a\mu}({\bf k})e^{i{\bf k}\cdot {\bf x}}
        + a^{\dagger}_{a\mu}({\bf k})e^{-i{\bf k}\cdot {\bf x}} \Big]\,, \\
&& \Pi_{a\mu}({\bf x}) = \int \frac{id^3k}{\sqrt{(2\pi)^3}}
             \sqrt{\frac{\Omega}{2}}
        \Big[ a_{a\mu}({\bf k})e^{i{\bf k}\cdot {\bf x}}
        - a^{\dagger}_{a\mu}({\bf k})e^{-i{\bf k}\cdot {\bf x}} \Big]\,, \\
&& \omega^{(1)}_a({\bf x}) = \int \frac{d^3k}{\sqrt{2|{\bf k}|(2\pi)^3}}
    \Big[ \overline{C}_a({\bf k})e^{i{\bf k}\cdot {\bf x}}
    + \overline{C}_a^{\dagger}({\bf k})e^{-i{\bf k}\cdot {\bf x}} \Big]\,, \\
&& \Pi^{(1)}_a({\bf x}) = \int \frac{(-i)d^3k}{\sqrt{(2\pi)^3}}
             \sqrt{\frac{|{\bf k}|}{2}}
    \Big[ C_a({\bf k})e^{i{\bf k}\cdot {\bf x}}
    + C^{\dagger}_a({\bf k})e^{-i{\bf k}\cdot {\bf x}} \Big]\,, \\
&& \omega^{(2)}_a({\bf x}) = \int \frac{(-i)d^3k}{\sqrt{2|{\bf k}|(2\pi)^3}}
    \Big[ C_a({\bf k})e^{i{\bf k}\cdot {\bf x}}
    - C^{\dagger}_a({\bf k})e^{-i{\bf k}\cdot {\bf x}} \Big]\,, \\
&& \Pi^{(2)}_a({\bf x}) = \int \frac{d^3k}{\sqrt{(2\pi)^3}}
             \sqrt{\frac{|{\bf k}|}{2}}
    \Big[\overline{C}_a({\bf k})e^{i{\bf k}\cdot {\bf x}}
    -\overline{C}^{\dagger}_a({\bf k})e^{-i{\bf k}\cdot {\bf x}} \Big]\,, \\
&& H_0 = -g^{\mu\nu} \int d^3k \Omega(|{\bf k}|)
         a^{\dagger}_{a\mu}({\bf k}) a_{a\nu}({\bf k})
         + \int d^3k |{\bf k}| \Big\{
         \overline{C}^{\dagger}_a({\bf k}) C_a({\bf k})
         + C^{\dagger}_a({\bf k}) \overline{C}_a({\bf k}) \Big\}
         + H^0_{\psi}\,,
\end{eqnarray}
where
\begin{eqnarray}
&& \Omega(|{\bf k}|) = \sqrt{M^2+|{\bf k}|^2} \,, \\
&& \Big[a_{a\mu}({\bf k}), a^{\dagger}_{b\nu}({\bf k}')\Big]
    = - g_{\mu\nu} \delta_{ab}\delta^3({\bf k}-{\bf k}') \,, \\
&&  \Big[ a_{a\mu}({\bf k}), a_{b\nu}({\bf k}')\Big] =
\Big[a^{\dagger}_{a\mu}({\bf k}), a^{\dagger}_{b\nu}({\bf k}')\Big] = 0 \,,\\
&& \Big[\overline{C}_a({\bf k}), C^{\dagger}_b({\bf k}')\Big]_+
   = \Big[C_a({\bf k}), \overline{C}^{\dagger}_b({\bf k}')\Big]_+
   = \delta_{ab}\delta^3({\bf k}-{\bf k}') \,, \\
&& \Big[C_a({\bf k}), C_b({\bf k}')\Big]_+
  = \Big[\overline{C}_a({\bf k}), \overline{C}_b({\bf k}')\Big]_+
  = \Big[C_a({\bf k}), \overline{C}_b({\bf k}')\Big]_+ \nonumber\\
&&\ \ \ \ \ \ \ \ \ \ \ \ \ \ \ \ \ \ \ \ \ \ \
  = \Big[\overline{C}_a({\bf k}), \overline{C}^{\dagger}_b({\bf k}')\Big]_+
  = \Big[{C}_a({\bf k}), C^{\dagger}_b({\bf k}')\Big]_+ = 0 \,.
\end{eqnarray}
The restriction of the Lorentz condition on a physical state
$|\Psi_{ph}\rangle$ can be expressed as
\begin{eqnarray}
&& k^{\mu} a_{a\mu}({\bf k}) |\Psi_{ph}\rangle = 0 \ \ \ \ \ \
   (k^0=\Omega(|{\bf k}|))  \,.
\end{eqnarray}
For a single particle state with the polarization vector
$\varepsilon^{\mu}({\bf k})$
\begin{eqnarray}
&& \varepsilon^{\mu}({\bf k}, \lambda) a^{\dagger}_{a\mu}({\bf k}) |0 \rangle
\end{eqnarray}
Eq. (17) gives
\begin{eqnarray}
&& k_{\mu}\varepsilon^{\mu}({\bf k}, \lambda) = 0 \,.
\end{eqnarray}
According to this condition two transversal polarization states
($\lambda=1,2$) and a longitudinal polarization state
($\lambda=3$) are allowed to be present for each {\bf k}.
\par
    Since $\partial^{\mu}\partial_{\mu}\omega_a(x)=0$ the theory we are
treating is invariant under the infinitesimal transformation
\begin{eqnarray}
&&
\delta A_{a\mu}(x) = \delta \zeta \partial_{\mu} \omega_a(x) \,, \ \ \ \
\delta \omega_a(x) = 0\,,\ \ \ \  \delta \overline{\omega}_a(x) = 0\,,\ \ \ \
\delta \psi(x) = 0 \,,
\end{eqnarray}
where $\delta \zeta$ is an infinitesimal fermionic real constant. Similarly,
since $\partial^{\mu}\partial_{\mu}\omega_a(x) = 0$ the theory is also
invariant under the transformation
\begin{eqnarray}
&& \delta A_{a\mu}(x) = \delta \zeta \partial_{\mu} \overline{\omega}_a(x) \,,
   \ \ \ \
\delta \omega_a(x) = 0\,,\ \ \ \  \delta \overline{\omega}_a(x) = 0\,,\ \ \ \
\delta \psi(x) = 0 \,.
\end{eqnarray}
This is the residual symmetry mentioned above. Under such a kind of
transformations a physical state should become an equivalent state.
Particularly
$\varepsilon^{\mu}({\bf k},\lambda)\delta a^{\dagger}_{a\mu}({\bf k})
|0 \rangle$ should be equivalent to zero, where
$\delta a^{\dagger}_{a\mu}({\bf k})$ is determined by
\begin{eqnarray}
&& \delta A_{a\mu}({\bf x})
        = \delta \zeta \big[\partial_{\mu}\omega_a\big]_{t=0} \,,
\end{eqnarray}
or by
\begin{eqnarray}
&& \delta A_{a\mu}({\bf x})
      = \delta \zeta \big[\partial_{\mu}\overline{\omega}_a\big]_{t=0} \,.
\end{eqnarray}
From (22), one has
\begin{eqnarray}
&& \delta a^{\dagger}_{a\mu}({\bf k})
= i \delta \zeta \widetilde{k}_{\mu} \sqrt{\Omega/|{\bf k}|}
    C^{\dagger}_a({\bf k})\,.
\end{eqnarray}
where the components of $\widetilde{k}_{\mu}$ are
\begin{eqnarray}
&& \widetilde{k}_0 = |{\bf k}|\,,\ \ \ \ \ \ \widetilde{k}_l = k_l \,.
\end{eqnarray}
Similarly from (23), one has
\begin{eqnarray}
&& \delta a^{\dagger}_{a\mu}({\bf k})
= i \delta \zeta \widetilde{k}_{\mu} \sqrt{\Omega/|{\bf k}|}
    \overline{C}^{\dagger}_a({\bf k})\,.
\end{eqnarray}
It follows that  $C^{\dagger}_a({\bf k}) |0 \rangle $  and
$\overline{C}^{\dagger}_a({\bf k})|0\rangle$ should be equivalent to zero.
\par
   In conclusion, the initial-final state subspace is formed by all the
physical states. A physical state can only contain material particles,
transversal polarization or longitudinal polarization gauge Bosons and does
not contain ghost particles.
\par
    We are grateful to Professor Yang Li-ming for helpful discussions. This
work was supported in part by National Natural Science Foundation
of China (19875002) and supported in part by Doctoral Program
Foundation of the Institution of Higher Education of China.
\par
\ \par
\begin{center}
{\large \bf References}
\end{center}
\par  \noindent
[1]\ Ze-sen Yang, Zhining Zhou, Yushu Zhong and Xianhui Li, hep-th/9912046.
\par  \noindent
[2]\ Ze-sen Yang, Xianhui Li, Zhining Zhou and Yushu Zhong, hep-th/9912034.
\par
\end{document}